\documentclass[pra,twocolumn,showpacs,showkeys,linenumbers]{revtex4}
\usepackage{amssymb,amsfonts,amsmath,enumerate}
\usepackage{graphicx}

\begin{document}

\title{Collision entropy and optimal uncertainty}

\author{G.M.\ Bosyk, M.\ Portesi and A.\ Plastino}
\affiliation{Instituto de F\'{\i}sica La Plata (IFLP, CONICET), and Dpto.\ de F\'{\i}sica, Fac.\ de Ciencias Exactas, Universidad Nacional de La Plata, 115 y 49, C.C.~67, 1900 La Plata, Argentina}

\begin{abstract}

We propose an alternative measure of quantum uncertainty for pairs of arbitrary observables in the 2-dimensional case, in terms of collision entropies. We derive the optimal lower bound for this entropic uncertainty relation, which results in an analytic function of the overlap of the corresponding eigenbases. Besides,  we obtain the minimum uncertainty states. We compare our relation with other formulations of the uncertainty principle.


\end{abstract}

\date{\today}

\maketitle

\section{Introduction}

Quantum mechanics' uncertainty principle (UP) is a fundamental
theoretical notion, being not just a side result of quantum
mechanics but arguably one of its most important fundamental
concepts. It establishes the existence of an irreducible lower
bound for the uncertainty in  preparing a system's state. The
original statement made use of the dispersion in an observable's
measurement. The concept of entropy \cite{wehrl} provided a
totally new perspective on UPs. So-called entropic uncertainty
relations (EURs) \cite{Birula2011,Wehner2010} are a relatively
recent, related concept, that greatly improves the original one in
the sense of allowing for nontrivial {\it state-independent} lower
bounds. The formulation of the UP in terms of EURs, besides being more
appropriate than the original statement from a theoretical point of view,
acquires significant importance in the quantum information theory realm \cite{Birula2011,Wehner2010}.
In particular, EURs provide entanglement criteria and foundations for
the security of many quantum cryptographic protocols, amongst other applications
\cite{Huang2011,Saboia2011,Zander2010,Walborn2009,Guhne2004,Giovannetti2004,Wu2009,Ballester2007,Renes2009,Berta2010,Tomamichel2011,Guerrero2011,Zozor2011,Honarasa2009}.

In this communication we are concerned with the study of uncertainty relations between two quantum observables in the
case of $2$-dimensional systems.
We will show that a  construct called collision
entropy, i.e.\ the R\'enyi entropy of index 2, exhibits
significant advantages as an uncertainty measure in this case. The paper is organized as follows. An abridged history of UP formulations can be found in Sect.~\ref{historic}, focusing on the concepts that we will develop further.
In Sect.~\ref{sect:optimalbound}, we consider the sum of collision entropies as UP-indicator for a pair of arbitrary observables in
2-dimensional Hilbert space. We obtain the optimal lower bound and the minimum uncertainty states for the proposed uncertainty measure, following the procedure by Ghirardi {\it et al.} \cite{Ghirardi2003} who exploit the Bloch
representation to facilitate the associated minimization problem's
tractability. Sect.~\ref{sect:otherrelations} is devoted to compare our
entropic uncertainty relation with other inequalities found in the
relevant literature. Finally, some conclusions are drawn in
Sect.~\ref{sect:conclusions}, stressing that the use of the collision entropy, compared to Shannon one, allows us to provide an analytical expression for the lower bound as well as the minimizing states.

\section{Some Historic considerations}
\label{historic}

\subsection{Entropy as an alternative uncertainty measure}

The original quantitative UP-formulation was proposed in the famous paper by Heisenberg \cite{Heisenberg1927} and demonstrated by Kennard \cite{Kennard1927} in the case of position and momentum observables. Robertson \cite{Robertson1929} extended the relation for other cases. These formulations were based on products of variances for pairs of observables.
A completely novel perspective for the UP, in the case of canonically conjugate variables, was introduced in the pioneering contributions by Hirschman \cite{Hirschman1957}, Mamojka \cite{Mamojka1974}, and Bialynicki-Birula and Mycielski \cite{Birula1975}, in the framework of information theory. Specifically, the new proposal was to employ the sum of Shannon entropies associated to the position and momentum distributions. In Ref.~\cite{Birula1975} it has been shown that the entropic relation is stronger than the Heisenberg one.

The introduction of the information-theory alternative was inspired by the power of entropy to describe properties such us uncertainty in connection to probability distributions. Indeed, information entropies become  much more appropriate in the quantum, probabilistic world. Additionally, as discussed in \cite{Birula2011}, the use of standard deviations to express indeterminacy has some limitations.
The original variance-formulation of UP has also been criticized~\cite{Deutsch1983,Uffink1988} on the grounds that the associated bound, given by the expectation value of the commutator between the two observables, depends (if the commutator is not a $c$-number) on the state of the system and thus lacks a universal character.
Moreover, it can be easily seen~\cite{Ghirardi2003} that for bounded operators the lower bound is trivially zero, yielding no valuable information.
An alternative that has been envisaged to circumvent this obstacle precisely consists in using as an UP-measure
the Shannon or other generalized entropies associated to the probability distributions of the two observables' outcomes.

De Vicente and Sanchez-Ruiz~\cite{deVicente2008} provided  the
best lower bound for the sum of Shannon entropies in the case of
an arbitrary pair of observables with discrete, $N$-level spectra
(see also Ref.~\cite{nos2011}). This was an improvement on the
Maassen-Uffink uncertainty relation presented in
Ref.~\cite{Uffink1988}, based on the Landau-Pollak
inequality~\cite{Landau1961}. In the particular case of single
qubit systems ($N=2$), Sanchez-Ruiz~\cite{Sanchez1998} and
Ghirardi {\it et al.}~\cite{Ghirardi2003}, independently extracted
the {\it optimal} lower bound for the Shannon entropies' sum. For
$N$ up to 5, a very recent study by Jafarpour and
Sabour~\cite{Jafarpour2011} gave (numerically) a more stringent
bound.

\subsection{Our entropic quantifier}

Now, the R\'enyi entropies \cite{Renyi1970} constitute a family of
generalized information-theoretic measures that account for the uncertainty or
lack of information associated to a probability distribution. In
the finite-dimensional, discrete case the  definition reads as
\begin{equation}
H_{q}\left( \left\{ p_{i}\right\} \right) =\frac{1}{1-q}\ln
\left(\sum_{i=1}^{N}p_{i}^{q}  \right),
\label{Renyi}
\end{equation}
where $0\leq p_{i}\leq 1$ and $\sum_{i=1}^{N}p_{i}=1$. $N$ is the
number of levels, and the index $q>0$ with $q\neq 1$. When
$q\rightarrow 1$, $H_q$ approaches the Shannon entropy $H_1\equiv
H=-\sum p_i \ln(p_i)$. In the particular case $q=2$ the R\'enyi
entropy is known as the collision entropy. This quantity is widely
used in quantum information process and quantum cryptography. The
collision entropy can be written in terms of the so-called  purity
of a given probability distribution, indeed $H_{2}\left( \left\{
p_{i}\right\} \right)$ is the natural logarithm of the inverse of
the purity, which is given by $\sum_{i=1}^{N}p_{i}^{2}$. A
particularly interesting scenario arises  when the entropic index
$q$ tends to infinity. Here the R\'enyi entropy, known as
min-entropy, becomes $H_\infty\left( \left\{ p_{i}\right\}
\right)=-\ln P$, where $P=\max_i\left( p_{i} \right)$.

EURs using
R\'enyi entropies as measures of uncertainty have been recently
studied in the
literature~\cite{Birula2006,Zozor2007,Zozor2008,Rastegin2010,Luis2011}.
However, most of the concomitant  EURs in these references deal
just with (i) {\it complementary} observables (i.e., those whose
eigenstates are linked by a Fourier transformation) and/or with
(ii) {\it conjugated} R\'enyi indices $q$ and $q'$ (i.e., when $\frac
1q+\frac 1{q'}=2$, which includes the Shannon case).

{\it The present contribution deals precisely with a novel
situation, not accounted for previously in related materials}. The
novelty resides in the fact  that we will not be restricted here
in the way mentioned above. Instead,  (i) arbitrary pairs of
observables will be considered and, (ii) we will adopt the
collision entropy as the uncertainty quantifier. In other words,
we propose and analyze an entropic uncertainty relation which
{\sf does not make use of the Riesz' theorem hypothesis of index
conjugation}.

\section{Derivation of the optimal bound for the sum of collision entropies}
\label{sect:optimalbound}

\subsection{Our optimal relation in terms of collision entropies}

The sum of the collision entropies for two observables
$A,B\in \mathbb{C}^{2\times 2}$
for a system prepared in the quantum pure state $\left\vert \Psi
\right\rangle \in \mathbb{C}^{2}$ is given by
\begin{eqnarray}
\lefteqn{\mathcal{U}(A,B;\Psi) \equiv H_{2}(A)+H_{2}(B)=} \nonumber \\
&& = - \ln \left( p_{1}^{2}\left(
A\right) +p_{2}^{2}\left( A\right) \right) -\ln \left( p_{1}^{2}\left(
B\right) +p_{2}^{2}\left( B\right) \right) ,
\label{U(A,B)}
\end{eqnarray}
where $p_{i}\left( A\right) =\left\vert \left\langle a_{i}\right. \left\vert
\Psi \right\rangle \right\vert ^{2}$ and $p_{i}\left( B\right) =\left\vert
\left\langle b_{i}\right. \left\vert \Psi \right\rangle \right\vert ^{2}$
are the probabilities for the outcomes of observables $A$ and $B$, respectively, whose
eigenbasis are denoted by $\{\left\vert a_{1}\right\rangle, \left\vert a_{2}\right\rangle\}$ and $\{\left\vert b_{1}\right\rangle, \left\vert b_{2}\right\rangle \}$,
respectively.

The minimization problem is significantly ameliorated if one exploits the well
known Bloch representation, along lines similar to those of Ref.~\cite{Ghirardi2003} that deals with Shannon-UP.
The most general normalized quantum pure state of a single qubit can be written (up to an unobservable phase factor) as \ $\left\vert \Psi \right\rangle =\cos \frac{\theta }{2}\left\vert
0\right\rangle +e^{i\phi }\sin \frac{\theta }{2}\left\vert 1\right\rangle$,
with $0\leq \theta \leq \pi$ and $0\leq \phi \leq 2\pi$, where $\{\left\vert
0\right\rangle ,\left\vert 1\right\rangle \}$ is the so called computational
basis. To each pure state $\left\vert \Psi \right\rangle$ a unique
point on the Bloch sphere is assigned, represented by the unit vector
$\vec{s}=(\cos\phi \sin\theta, \sin\phi \sin\theta, \cos\theta) \in {\mathbb{R}^{3}}$.
In terms of this Bloch vector one can write down the density operator associated to the (pure) state of the system:
 \ $\rho=|\Psi\rangle\langle\Psi|=\frac 12(I+\vec s\cdot\vec\sigma)$, with $\vec{\sigma }=\left( \sigma _{X},\sigma _{Y},\sigma
_{Z}\right)$ denoting the Pauli matrices and $I$ the $2\times 2$ identity matrix.
The observables in this representation acquire the form
\begin{align}
A& =\alpha _{1}I+\alpha _{2}\,\vec{a}\cdot \vec{\sigma }
\label{A} \\
B& =\beta _{1}I+\beta _{2}\,\vec{b}\cdot \vec{\sigma }
\label{B}
\end{align}
where $\vec{a}, \vec{b}\in {\mathbb{R}^{3}}$ are unit vectors and $\alpha_{1}, \alpha_{2}, \beta_{1}$, and $\beta_{2}$ are real parameters.

We want the tightest lower bound for the uncertainty measure~\eqref{U(A,B)} for given observables $A$ and $B$, over all possible states $|\Psi\rangle$. This is equivalent to searching for
\begin{equation}
\min_{\theta ,\phi }\,\mathcal{U}(A,B;\Psi ) ,
\label{minU}
\end{equation}
for $\alpha_{i}$, $\beta_{i}$, $\vec{a}$, and $\vec{b}$ fixed.
Without loss of generality, we consider $\vec{a}\cdot
\vec{\sigma }$ instead of $A$ (they have the same
eigenbasis and then the same R\'enyi entropy). Analogously, we consider $\vec{b}\cdot \vec{\sigma }$ instead of $B$.
Let us pass to the squared moduli of the inner products of eigenstates of $A$ and $B$.
In terms of the scalar product between the corresponding unit vectors $\vec{a}$ and $\vec{b}$ we have
\begin{equation}
\left(\,\left\vert \left\langle a_{i}\right. \left\vert b_{j}\right\rangle
\right\vert ^{2}\,\right)=\left(
\begin{array}{cc}
\frac{1+\vec{a}\cdot \vec{b}\,}{2} & \frac{1-\vec{a}\cdot \vec{b}\,}{2} \\
\frac{1-\vec{a}\cdot \vec{b}\,}{2} & \frac{1+\vec{a}\cdot \vec{b}\,}{2}
\end{array}
\right) .
\label{matrix}
\end{equation}
The greatest element is known as the overlap between eigenbases.
Hence, denoting by $\gamma$ the angle formed by the $\vec{a}$ and $\vec{b}$ directions, the overlap becomes
\begin{eqnarray}
c \equiv \max_{i,j=1,2}\left\vert \left\langle a_{i}\right. \left\vert
b_{j}\right\rangle \right\vert &=&
\max_{\gamma \in (0,\pi)}\{\cos
\frac{\gamma}{2},\sin \frac{\gamma}{2}\}= \nonumber \\
 &=& \left\{
\begin{array}{c}
\cos \frac{\gamma}{2}\quad \text{ if }0< \gamma \leq \pi /2 \\
\sin \frac{\gamma}{2}\quad \text{ if }\pi /2\leq \gamma < \pi,
\end{array}
\right.
\label{overlap}
\end{eqnarray}
where we restrict the values of $\gamma$ to the interval $(0,\pi)$. Due
to symmetry arguments, results for $\gamma\in(\pi,2\pi)$ can be obtained straightforwardly. Besides, $\gamma=0$ and $\gamma=\pi$ (implying $c=1$) are excluded since they correspond indeed to pairs of commuting observables.
The particular case $\gamma=\pi/2$ gives $\left\vert \left\langle a_{i}\right. \left\vert b_{j}\right\rangle
\right\vert =1/\sqrt{2}$ \ for all $i,j=1,2$, corresponding to that special situation in which
 the observables are complementary. For the two-dimensional case, the range for the overlap $c$ goes from $1/\sqrt 2$ up to~$1$.

As our main result, we will show that the
uncertainty measure~\eqref{U(A,B)} has as a lower bound a function
depending only on the overlap $c\in[1/\sqrt 2,1)$, of the form
\begin{equation}
\mathcal{U}(A,B;\Psi )\equiv H_{2}(A)+H_{2}(B) \geq
-2\ln \frac{1+c^{2}}{2} .
\label{EUR}
\end{equation}
Moreover the uncertainty measure~\eqref{U(A,B)} exhibits an upper
bound, since $\mathcal{U}(A,B;\Psi)\leq 2\ln 2$.  {\it We stress
that the EUR~\eqref{EUR} is valid for arbitrary pairs of
(2-dimensional) observables, not merely for those special ones
that are complementary.}

\subsection{Derivation of the optimal bound and minimal uncertainty states}

In order to demonstrate the above result, let us first write down the uncertainty measure $\mathcal{U} (A,B;\Psi)$ in terms of the scalar products of $\vec{a}$ and $\vec{b}$ with the Bloch vector $\vec{s}$. A short calculation yields
\begin{equation}
\mathcal{U}=\mathcal{U}(\vec a,\vec b;\vec s)=-\ln \frac{1+( \vec{a}\cdot
\vec{s}) ^{2}}{2}-\ln \frac{1+( \vec{b} \cdot \vec{s}) ^{2}}{2} .
\label{U1}
\end{equation}
Therefore, the extremization of $\mathcal{U}$ becomes a geometric
problem: for fixed directions $\vec{a}$ and $\vec{b}$, we need to find the unit vector $\vec{s}$ that bounds either from below or from above the quantity~\eqref{U1}. Trivially, the maximum of this quantity corresponds to the case when $\vec s$ is just one of the two unit vectors along the direction perpendicular to both $\vec a$ and $\vec b$. Then $\mathcal{U}_\mathrm{Max}=2 \ln 2$. This happens indeed when all $p_i=1/2$ and then the collision entropy for each
observable is separately maximal.
Let us now show that the minimum of the quantity~\eqref{U1} is given when $\vec{a}$, $\vec{b}$ and $\vec{s}$ are coplanar, a fact that reduces
the number of variables in the minimization problem. Consider the function $U(x)=-\ln \left(\frac{1+x^{2}}{2}\right)$ with $x\in \lbrack 0,1]$.
Straightforwardly, one sees that $U(x)$ is a strictly decreasing function in its domain. Thus, for a value $x_{0}\in \lbrack 0,1]$ one has $U(x)\geq U(x_{0})$
for all $x\leq x_{0}$. Let $\Pi$
be the plane determined by the fixed vectors $\vec{a}$ and $\vec{b}$, and let $\vec{d}$ be
any unit vector belonging to an arbitrary orthogonal plane $\Pi^{\bot}$.
If $\vec{d_{0}}$ is one of the two unit vectors that belong to $\Pi$ and $\Pi^{\bot}$, then $\vert \vec{a}\cdot
\vec{d}\vert \leq \vert \vec{a}\cdot
\vec{d_{0}}\vert$ for all $\vec{d}\in \Pi^{\bot}$. Thus, $U(\vert \vec{a}\cdot
\vec{d}\vert )\geq U(\vert\vec{a}\cdot \vec{d_{0}}\vert )$, the
equality being satisfied when $\vec{d}=\pm \vec{d_{0}}$,
i.e., when $\vec{d}\in\Pi$ as well. An analogous result is obtained
by changing $\vec{a}$ for $\vec{b}$. This justifies the fact that the
minimum of $\mathcal{U}$ will be reached under the condition that $\vec{a}$, $\vec{b}$, and $\vec{s}$ all belong to the same plane.
However, we still need to determine the direction of $\vec s$ relative to the fixed vectors $\vec a$ and $\vec b$.

Let us denote by $\chi$ the angle between $\vec{a}$ and $\vec{s}$. Accordingly the uncertainty measure~\eqref{U1},
expressed in terms of the angles $\chi$ and $\gamma$, becomes
\begin{equation}
\mathcal{U}_{\gamma}(\chi )=-\ln \left(\frac{1+\cos ^{2}\chi }{2}\right)-\ln \left(\frac{1+\cos ^{2}\left( \gamma -\chi \right) }{2}\right) ,
\label{U}
\end{equation}
where $\chi$ can be restricted to the interval $[0,\pi]$ due to the
periodicity of this function. Thus, the minimization problem~\eqref{minU} reduces to that of finding the minimum of $\mathcal{U}_{\gamma}(\chi)$ for $\gamma$
fixed. Equating to zero the first derivative of $\mathcal{U}_{\gamma}(\chi)$ with respect to $\chi$ we arrive at the condition for a critical point, in the fashion
\begin{equation}
f\left( \chi \right) =f\left( \gamma -\chi \right) ,
\label{DU}
\end{equation}
where we have defined $f\left( x\right) =\frac{\sin 2x}{3+\cos 2x}$.

Let us solve now Eq.~\eqref{DU}. First, for any
fixed $\gamma$ we have the trivial solutions $\chi_{k}=\frac{\gamma +k\pi}{2}$ with $k\in{\mathbb{Z}}$. Thus, in the interval $\chi\in \lbrack 0,\pi]$ the two trivial solutions are
\begin{align}
\chi_<(\gamma)\equiv\chi _{1}& =\frac{\gamma}{2} \label{chi1}, \\
\chi_>(\gamma)\equiv\chi _{2}& =\frac{\gamma}{2} +\frac{\pi}{2} . \label{chi2}
\end{align}
These solutions correspond to the straight lines plotted in Figure~1.
Geometrically, the solution $\chi_{<}$ corresponds to the vector $\vec b+\vec a$, pointing in the direction of the interior bisector of the
angle determined by $\vec{a}$ and $\vec{b}$ directions. The solution $\chi_{>}$ corresponds to
a different vector, pointing along the direction of $\vec b-\vec a$ and being perpendicular to the former.
The norms of these two vectors can be simply expressed in terms of the overlap, since $\|\vec b\pm\vec a\|=2 c$.
Therefore, the Bloch vectors corresponding to the solutions~\eqref{chi1} and~\eqref{chi2} become: \ $\vec s_{\lessgtr}=\frac{\vec b\pm\vec a}{2c}$.
Notably, these solutions have also been obtained by Ghirardi {\it et al.} in Ref.~\cite{Ghirardi2003}, where Shannon entropies have been employed instead.

In addition to the trivial solutions, for a given range of $\gamma\in \lbrack \gamma^*,\gamma^{**}]$ it can be seen that there exist other solutions to Eq.~\eqref{DU}.
Unfortunately, they do not possess analytical expressions and have
to be calculated numerically. The limiting (critical) values of $\gamma^*$ and $\gamma^{**}$ for the existence of two or more than two solutions satisfy
\begin{equation*}
f^{\prime }\left( \chi \right) =f^{\prime }\left( \gamma -\chi \right) ,
\end{equation*}
coming from the condition that the maxima or, respectively, the minima of $f(\chi)$ and $f(\gamma-\chi)$ coincide. We can obtain in analytical fashion these critical values:
\begin{equation}
\gamma^*=\pi-\arccos \left( -1/3\right),
\; \text{ and } \;
\gamma^{**}=\arccos \left( -1/3\right) .
\end{equation}

\begin{figure}[htbp]
\centerline{\includegraphics[scale=0.75]
 {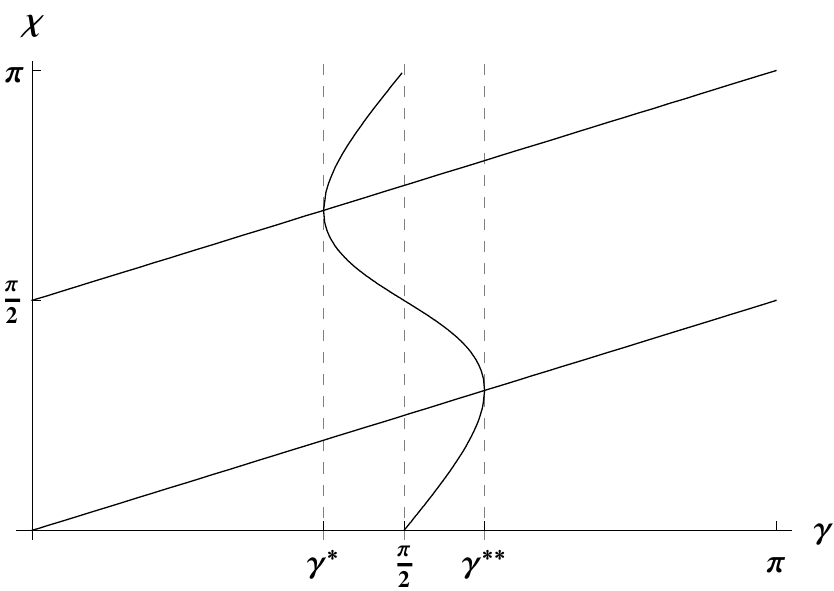}}
\caption{Solutions of Eq.~\eqref{DU}, for the angle $\chi$ between the Bloch vector $\vec s$ and the vector $\vec a$ corresponding to observable $A$. Operator $B$ enters through the parameter $\gamma$ which stands for the angle between $\vec a$ and $\vec b$. }
\label{figchivsgamma}
\end{figure}

We plot in Fig.~\ref{figchivsgamma} all solutions for $\chi$ in terms of the parameter $\gamma$.
We see that $\gamma^*$, $\pi/2$ and $\gamma^{**}$ are critical parameters, in the sense that the
number of solutions of Eq.~\eqref{DU} changes.
Let us now discuss in some detail the solutions pertaining to different regions of
the parameter~$\gamma$.
\begin{enumerate}
\item\label{case1} $\gamma \in (0,\gamma^*]$: the only solutions of Eq.~\eqref{DU} are
the trivial ones $\chi_{<}$ and $\chi_{>}$ in Eqs.~\eqref{chi1} and~\eqref{chi2}. Note that $\chi_{<}$ corresponds to the minimum of $\mathcal{U}_{\gamma}$ and $\chi_{>}$ to the maximum.
\item\label{case2} $\gamma \in (\gamma^*,\pi/2)$: $\chi_{<}$ and $\chi_{>}$ are still solutions of Eq.~\eqref{DU} but there exist other
two solutions, whose values can be calculated numerically for each $\gamma$.
The solution $\chi_{<}$ remains the absolute minimum of $\mathcal{U}_{\gamma}$ while $\chi_{>}$ is now a relative minimum and the
other solutions yield the (same) maximum.
\item\label{case3} $\gamma=\pi/2$: this is a special case because
the observables $A$ and $B$ are {\it complementary}, i.e., the overlap is precisely $c=1/\sqrt{2}$. The solutions $\chi_{<}=\pi/4$ and $\chi_{>}=3\pi/4$
yield the same minimum uncertainty measure,
while the other three ones (0, $\pi/2$, and $\pi$) give the maximum.
\item $\gamma \in (\pi/2,\gamma^{**})$: in this interval $\chi_{<}$
and $\chi_{>}$ reverse their behaviors with respect to those of Point~\ref{case2}.
That is, $\chi_{<}$ happens to be a relative minimum while $\chi_{>}$ is the absolute
minimum of $\mathcal{U}_{\gamma}$. The two other solutions correspond to maxima.
\item\label{case5} $\gamma \in [\gamma^{**},\pi)$: now $\chi_{<}$ corresponds to the maximum of $\mathcal{U}_{\gamma}$ and $\chi_{>}$ to the minimum. No other solutions exist in this interval.
\end{enumerate}
Thus, depending on $\gamma$, the value of $\chi$ at which $\mathcal{U}_{\gamma}$ acquires its minimum can be given in
 concise fashion using the Heaviside function as
\begin{equation}
\chi_{\min}(\gamma)= \frac{\gamma}{2}+\frac{\pi}{2} \, \Theta\left(\gamma-\frac{\pi}{2} \right) , \qquad \gamma \in (0,\frac{\pi}{2})\cup(\frac{\pi}{2},\pi) ,
\end{equation}
while $\chi_{\min}$ at $\gamma=\pi/2$ takes the values $\pi/4$ and $3\pi/4$ as discussed in Point~\ref{case3}.
Summing up, replacing the values of $\chi_{<}$ and $\chi_{>}$ in the uncertainty measure~\eqref{U} we obtain its minimum as a function of~$\gamma$, and finally
\begin{eqnarray}
\mathcal{U}(A,B;\Psi) &\geq&
\mathcal{U}_{\min }(A,B)= \nonumber \\
&=& \! \left\{
\begin{array}{c}
\! -2\ln \left(\frac{1+\cos ^{2}\frac{\gamma}{2}}{2}\right) \; \! \text{ if } 0< \gamma \leq
\pi /2 \\
-2\ln \left(\frac{1+\sin ^{2}\frac{\gamma}{2}}{2}\right) \; \text{ if } \pi /2\leq \gamma < \pi .
\end{array}
\right.
\label{Umin}
\end{eqnarray}
Recalling Eq.~\eqref{overlap}, we complete the proof of the UP-relation proposed in~\eqref{EUR}.

For the sake of completeness {\it we also determine the states
saturating our UP-relation}. As already mentioned, these states
have Bloch vectors $\vec s_<$ or $\vec s_>$ depending whether
$\gamma$ is, respectively, smaller or larger than $\pi/2$.
Therefore, the corresponding minimum-uncertainty density operators
become
\begin{eqnarray}
\rho_<=\frac 12 \left(I+\frac{\vec b+\vec a}{2\cos\frac{\gamma}{2}} \cdot\vec{\sigma}\right) & \quad & \text{ if }0< \gamma \leq
\pi /2,
\nonumber \\
\rho_>=\frac 12 \left(I+\frac{\vec b-\vec a}{2\sin\frac{\gamma}{2}} \cdot\vec{\sigma}\right) & \quad & \text{ if }\pi /2\leq \gamma < \pi .
\end{eqnarray}
Notice that for each $\gamma$ there exists another minimum-UP state arising
from the $\chi$-minimization of the uncertainty measure~\eqref{U} in the interval $\chi\in[\pi, 2\pi]$, that we did not consider due to the periodicity of this function. The solutions are $\tilde\chi_<=\frac{\gamma}{2}+\pi$ and $\tilde\chi_>=\frac{\gamma}{2}+\frac 32 \pi$. The discussion of Points \ref{case1} to \ref{case5} above apply also, replacing $\chi$ by $\tilde\chi$.
It is not difficult to see that the optimum states $\tilde\rho_{\lessgtr}$ have Bloch vectors
$\vec{\tilde s}_{\lessgtr}=-\,\frac{\vec b\pm\vec a}{2c}$ for $\gamma\lessgtr\pi/2$, respectively.

The particular case $\gamma=\pi/2$ (complementary observables) deserves some comment. In this case, there exist {\it four} states that minimize the uncertainty measure. To fix ideas, consider as an example the pair of observables $A=\sigma_x$ and $B=\sigma_y$. Their Bloch representations correspond to $\vec a=\breve{\i}$ and $\vec b=\breve{\j}$, respectively, so that we speak of $\gamma=\pi/2$ and $c=1/\sqrt 2$. Our approach prescribes that any state of the
system will have a collision-entropy uncertainty greater than or equal to $2\ln(4/3)$,
with equality for the states $|\Psi_l\rangle=\frac{1}{\sqrt 2}\left( |0\rangle+i^{l+1/2}|1\rangle\right)$, for $l=0,1,2$ and 3, up to a global phase factor.

We can go beyond the case of complementary observables, since our inequality~\eqref{Umin} allows us to quantitatively study
 the uncertainty related to the measurement of {\it any} pair of 2D observables, i.e., for {\it any} value of the overlap
$c\in[1/\sqrt 2,1)$. {\it We emphasize that, up to our knowledge,
such a possibility has not been yet explored (exhaustively) in the
literature.} For instance, among other interesting situations, we
can deal with the Hadamard gate and the $x$ (or $z$)
spin-projection. Assume $A=\frac{\sigma_x+\sigma_z}{\sqrt{2}}$ and
$B=\sigma_z$. Then their corresponding vectors $\vec
a=(1,0,1)/\sqrt 2$ and $\vec b=(0,0,1)$ determine an angle
$\gamma=\pi/4$ and, consequently, $c=\cos(\pi/8)\approx 0.924$.
The following relation ensues: \ $2 \ln 2\geq
H_2\left(\frac{\sigma_x+\sigma_z}{\sqrt
2}\right)+H_2(\sigma_z)\geq 2\ln\left(\frac{8}{6+\sqrt
2}\right)\approx 0.152$, with saturation of the last inequality
for those qubit-states that lie on the $xz$-plane ($\phi=0$) with
$\theta=\pi/8$ or $\tilde\theta=9\pi/8$.

\section{Comparison with other uncertainty relations}
\label{sect:otherrelations}

We pass to discuss and compare our present results with other formulations of the uncertainty principle.

\subsection{Heisenberg-Robertson inequality in terms of standard deviations}

We begin with the celebrated Heisenberg-Robertson (HR) inequality. For any
pair of arbitrary observables $A, B$ and a system described by the state $|\Psi\rangle$ one has
\begin{equation}
\Delta_{\Psi} A \ \Delta_{\Psi} B \geq \frac{1}{2}\left\vert
\left\langle \left[ A,B\right] \right\rangle_{\Psi} \right\vert ,  \label{H-R}
\end{equation}
where $\left\langle O\right\rangle_{\Psi} =\langle\Psi|O|\Psi\rangle$ is the mean
 value and $\Delta_{\Psi} O =\sqrt{\left\langle O^{2}\right\rangle_{\Psi} - \left\langle
O\right\rangle_{\Psi} ^{2}}$ the standard deviation of
the observable $O$. In the 2D case, using the Bloch representation, the HR inequality~\eqref{H-R} reads
\begin{equation}
|\alpha_{2}| \sqrt{1-(\vec{a} \cdot \vec{s})^{2}} \ |\beta_{2}| \sqrt{1-(\vec{b} \cdot \vec{s})^{2}} \geq
\left\vert \alpha _{2}\beta _{2}\right\vert \left\vert ( \vec{a}\times \vec{b}) \cdot
\vec{s}\right\vert .
\label{H-R2D}
\end{equation}
Some questions regarding~\eqref{H-R2D} may be cited here. If $|\Psi\rangle$ is an eigenstate of one of the two observables, $\vec s$ is parallel to the vector representing that observable and the HR-UP becomes trivial. No UP-information is gained thereby (in terms of a standard deviation) for measuring the other observable. Moreover, for {\it any} state whose Bloch vector $\vec{s}$ belongs to the plane determined
by $\vec{a}$ and $\vec{b}$, the right hand side of~\eqref{H-R2D} vanishes, thus
representing trivial information about the bound for the standard deviations' product: \ $\Delta A \, \Delta B\geq 0$.
Notice that in the derivation of our entropic uncertainty relation~\eqref{EUR} we showed that the minimum of $\mathcal{U}$ is attained for the case when $\vec{a}$, $\vec{b}$, and $\vec{s}$ lie in the same plane but only for those states satisfying Eqs.~\eqref{chi1} or~\eqref{chi2}. Furthermore we stress that, unlike HR-UP, the bound we obtain
is strictly greater than zero. Therefore, we conclude that, in 2D,
{\it relevant} information about the uncertainty of the observables can be obtained by recourse to the collision entropy.

\subsection{Luis relation in terms of purities}

In the appendix of Ref.~\cite{Luis2007} Luis derived, for {\it complementary} observables with discrete
spectrum of $N$ states, an uncertainty relation
following the work of Larsen~\cite{Larsen1990}. In his notation
\begin{equation}
\delta A\,\delta B\geq \left( \frac{2N}{N+1}\right) ^{2} ,
\label{Luis}
\end{equation}
where $\delta O=1/\sum_i p_i(O)^2$ is the inverse of the purity (or participation ratio) associated to observable $O$.
The expression~\eqref{Luis} is an improvement of the certainty relations that express the complementarity property, obtained by Luis in Refs.~\cite{Luis2001,Luis2003} for the case of $2$-dimensional systems and general $N$-dimensional systems, respectively.
Taking natural
logarithm in~\eqref{Luis}, this relation can be
expressed in terms of the sum of collision entropies: $H_{2}(A)+H_{2}(B)\geq 2\ln \frac{2N}{N+1}$. In the case $N=2$ (1-qubit system) that we are here considering, the Luis bound is $2\ln(4/3)$. This result coincides with our bound in~\eqref{EUR}
when $c=1/\sqrt{2}$. While we extend the 2D-formulation of the uncertainty principle to arbitrary observables (i.e., for any overlap $c$), Luis obtains instead results that are valid only for complementary observables, although for arbitrary (finite) $N$-dimensions.

\subsection{Landau-Pollak relation in terms of maximum probabilities}

The Landau-Pollak relation states that
\begin{equation}
\arccos \sqrt{P_{A}}+\arccos \sqrt{P_{B}}\geq \arccos c ,
\label{LP}
\end{equation}
where $P_{O}=\max_{i}p_{i}(O)$ for $O=A,B$. This inequality is another
alternative to the UP-mathematical formulation, introduced for time-frequency analysis in~\cite{Landau1961} and
adapted to physics two decades later~\cite{Uffink1988}. We have in this regards an interesting
result: the states that minimize $\mathcal{U}(A,B;\Psi )$ also saturate the inequality~\eqref{LP},
which means that the lower bound in the Landau-Pollak uncertainty
relation is optimal in 2D. Indeed, if we compute the maximum probabilities for the states that minimize $\mathcal{U}$, we find
\begin{equation}
P_{A}=P_{B}=\left\{
\begin{array}{c}
\cos ^{2}\frac{\gamma}{4} \quad \text{ if }0< \gamma \leq \pi /2 \\
\sin ^{2}\frac{\gamma +\pi }{4} \quad \text{ if }\pi /2\leq \gamma < \pi,
\end{array}
\right.
\label{Pgamma}
\end{equation}
in terms of the angle $\gamma$. Using Eq.~\eqref{overlap}, this can be simply reformulated as
\begin{equation}
P_{A}=P_{B}=\frac{1+c}{2},
\label{Pc}
\end{equation}
in terms of the overlap~$c$.
On the other hand, when~\eqref{LP} becomes an equality, the Landau-Pollak relation can be recast in the fashion
\begin{equation}
c=\sqrt{P_{A}P_{B}}-\sqrt{\left( 1-P_{A}\right) \left( 1-P_{B}\right) } .
\label{c}
\end{equation}
It is easy to see that replacing~\eqref{Pgamma}, or equivalently~\eqref{Pc}, in~\eqref{c}, the right hand side
of this equation identically yields $c$ for any $\gamma$. Therefore, the
Landau-Pollak relation~\eqref{LP} and the entropic uncertainty relation~\eqref{EUR}
(as well as its analogous Shannon-entropy expression~\cite{Sanchez1998,Ghirardi2003})
are equivalent for 1-qubit systems.

\subsection{Maassen-Uffink relation in terms of min-entropy}

In Ref.~\cite{Uffink1988}, Maassen and Uffink~(MU) work out an improvement of the Shannon-entropy uncertainty relation
 and advance an entropic UP for the sum of the min-entropies $H_{\infty}(O)=-\ln P_O$ associated to two observables $A$ and $B$
characterized by finite, discrete spectra. The pertinent expression reads
\begin{equation}
H_{\infty}(A)+H_{\infty}(B)\geq -2\ln \frac{1+c}{2} .
\label{MU}
\end{equation}
They encounter this relation by maximizing the product of the
maximum probabilities, $P_A P_B$, subject to the Landau-Pollak
inequality~\eqref{LP}. It is straightforwardly seen that,
replacing~\eqref{Pgamma} [or equivalently~\eqref{Pc}], in the left
hand side of~\eqref{MU}, we obtain an equality. Therefore, we find
that the states that saturate our EUR~\eqref{EUR} given in terms of collision entropies, also saturate the MU-EUR~\eqref{MU} that uses  min-entropies, with an equality in the Landau-Pollak
relation. One may be tempted to conjecture that in the $N=2$ case,
for any entropic index $q>0$, the minimum of the sum of the
$q$-R\'enyi entropies is reached when $P_{A}=P_{B}=\frac{1+c}{2}$.
Let us define the function $\mathcal{F}_q(c)\equiv \frac{2}{1-q}
\ln \left[\left(\frac{1+c}{2}\right)
^q+\left(\frac{1-c}{2}\right)^q \right]$. The question is whether
one can assure that $H_q(A)+H_q(B)\geq \mathcal{F}_q(c)$ for any
positive $q$. In the particular cases $q=2$ and
$q\rightarrow\infty$, respectively, we do obtain the lower bounds
$\mathcal{F}_2(c)=-2\ln \frac{1+c^2}{2}$ and
$\mathcal{F}_\infty(c)=-2\ln \frac{1+c}{2}$, that in turn
correspond to the right hand sides in the EURs~\eqref{EUR}
and~\eqref{MU}. However, we can not prove at this point that the
claim remains valid neither for any $q$ nor for any arbitrary pair
of 2D observables. As a counterexample, consider for instance the
$q\rightarrow 1$ Shannon case. It has been proved that the
function $\mathcal{F}_1(c)=-(1+c) \ln \frac{1+c}{2}-(1-c) \ln
\frac{1-c}{2}$  gives the absolute minimum of $H_1(A)+H_1(B)$ {\it
only} when the overlap $c$ belongs to the interval $[c^*,1)$, with
$c^*\simeq 0.834$ (we refer the reader to~\cite{nos2011}, where a
detailed analytical study of this point is provided). In other
words, for those pairs of 2D-observables with overlap between
$1/\sqrt 2$ and $c^*$, Eq.~\eqref{Pc} does not correspond to the
optimal solution regarding the minimization problem for the sum of
Shannon entropies.

\section{Concluding remarks}
\label{sect:conclusions}

In the 1-qubit scenario we have derived an optimal lower bound for the collision entropies' sum associated
to an arbitrary pair of observables. Although we have dealt with the simplest conceivable system, the
relevance of our entropic uncertainty relation given in~\eqref{EUR} is that:
\begin{itemize}
\item we obtain a lower bound that is optimal,
\item we find indeed the family of states that saturate the inequality,
\item we consider arbitrary pairs of observables, and
\item we take into account pairs of R\'enyi entropies where the
corresponding indices are not conjugate ones.
\end{itemize}
We emphasize that the conjunction of the last two points has not received much attention in the literature. Previous works were based on the Riesz theorem, which imposes the conjugacy restriction for the entropic indices.

Another advantage of using collision entropies, as compared with results given in terms of Shannon ones (see Refs.\ \cite{Sanchez1998,Ghirardi2003}), is that the lower bound in our case is analytical. This could be useful for future applications, for instance in connection with entanglement criteria, state discrimination, quantum cryptographic protocols, etc.

Moreover, we have shown that the states that minimize the collision entropy UP-measure defined by Eq.~\eqref{U(A,B)} also
saturate the EUR~\eqref{MU} given by Maassen and Uffink and,
additionally, saturate the Landau-Pollak relation~\eqref{LP}. They yield no relevant information concerning the Heisenberg-Robertson standard-deviation formulation, which turns out to be trivial in our scenario.

Furthermore,
it can be proved that the existence of relation~\eqref{MU} guarantees a non-trivial entropic uncertainty inequality for R\'enyi entropies of arbitrary (positive) indices. This is done making use of the monotonicity property of the family of R\'enyi entropies $H_q$ with respect to the index $q$.
The present study has allowed us to advance entropic UPs of the form
\begin{equation}
H_q(A)+H_{q'}(B) \geq -2\ln \frac{1+c^{2}}{2} ,
\label{EUR2}
\end{equation}
for {\it any} couple $(q;q')\in \mathcal{R}=\{0<q \leq 2$, $0<q' \leq 2\}$, where $A$ and $B$ are {\it any} arbitrary 2D-observables.
Within the region $\mathcal{R}$ of the $q$-$q'$ plane, the relation~\eqref{EUR2} is more stringent than the one derived following Maassen-Uffink's prescription~\eqref{MU}.
In order to prove the assertion~\eqref{EUR2} we just need the fact that R\'enyi entropy is strictly decreasing with the entropic index.
Thus, the l.h.s.\ in~\eqref{EUR2} becomes greater than or equal to $H_2(A)+H_2(B)$, which in turn is lower bounded as in~\eqref{EUR}.
The uncertainty relation~\eqref{EUR2} is in general non-optimal. We claim that at least it is optimal at the vertex $(q;q')=(2;2)$ of the rectangular region $\mathcal{R}$.

Note that the extension of our EUR to {\it mixed states} can be easily made due to the fact that the collision entropy is a concave
function for 1-qubit systems~\cite{Ben1978}.
Generalizations to $N$-level systems are the subject of active current research.

\section*{ACKNOWLEDGMENTS}

This work has been supported by PICT-2007-806 (ANPCyT) and PIP~1177/09 (CONICET), Argentina.

\end{document}